\documentclass[twocolumn,a4paper,10pt, bibliography=totocnumbered]{scrartcl}
\input{paper.sty}
\usepackage{calrsfs}
\usepackage{cuted}

\definecolor{KBFIred}{RGB}{163,35,47}

\newcommand{\mpl}{M_{\text{P}}}
\newcommand{\mfp}{m_{\text{FP}}}
\newcommand{\fmn}{f_{\mu \nu}}
\newcommand{\gmn}{g_{\mu \nu}}
\newcommand{\Gmn}{G_{\mu \nu}}
\newcommand{\Mmn}{M_{\mu \nu}}
\newcommand{\bfmn}{\bar{f}_{\mu \nu}}
\newcommand{\bgmn}{\bar{g}_{\mu \nu}}
\newcommand{\hmn}{h_{\mu \nu}}
\newcommand{\lmn}{\ell_{\mu \nu}}
\newcommand{\cGmn}{{\mathcal{G}}_{\mu \nu}}
\newcommand{\cG}{{\mathcal{G}}}
\newcommand{\Emnlk}{\mathcal{E}_{\mu\nu}^{\lambda\kappa}}
\newcommand{\D}{\nabla}
\newcommand{\Mij}{M_{ij}}

\usepackage{titlesec}
\titleformat*{\section}{\bfseries}
\titleformat*{\subsection}{\itshape}

\begin{document}

\title{\textcolor{KBFIred}{Oscillating Spin-2 Dark Matter
}\vspace{.5em}}

\author[]{Luca Marzola
}
\author[]{Martti Raidal
}
\author[]{Federico R. Urban
}

\affil[]{Laboratory of High Energy and Computational Physics, National Institute of Chemical Physics and Biophysics, R\"avala pst.~10, 10143 Tallinn, Estonia.}
\vspace{1cm}
\date{\normalsize{Dated: \today}}

\twocolumn[
\begin{@twocolumnfalse}
\maketitle

\begin{abstract}
	\noindent
{\textbf{Abstract:}}
We show that the coherent oscillations of a spin-2 field from bimetric theory can easily account for the observed dark matter abundance.  We obtain the equation of motion for the field in a cosmological setting and discuss in detail the phenomenology of the model. The framework is testable in precision measurements of oscillating electric charge in atomic clocks, using atomic spectroscopy and in dedicated resonant mass detectors as well as in axion-like-particles experiments, which therefore provide a new window to probe and test gravity itself.  We also comment on possible multimetric extensions of the framework that straightforwardly implement the clockwork mechanism for gravity.
\vspace{1cm}
\end{abstract}
\end{@twocolumnfalse}]


\section*{Introduction} 
\label{sec:Introduction}

The presence of dark matter (DM) in our Universe is inferred purely from its  gravitational effects~\cite{Jungman:1995df,Bertone:2004pz}, availing the idea that this mysterious component is a manifestation of gravity itself. This possibility, however, immediately faces two difficulties. First, the lack of a consistent quantum theory of gravity gives no {\itshape theoretical} guidance to move beyond the paradigm of General Relativity (GR).
Second, the characteristic weakness of gravitational interactions strongly limits the {\itshape experimental} guidance on which a new theory can be built.
Nonetheless, in this work we suggest that an approach driven by gravity is actually more rewarding.

DM is expected to arise in a natural way within theoretically consistent extensions of GR, as these theories necessarily include new gravitational degrees of freedom. The interactions of the latter are dictated by the theoretical framework itself, likewise for neutralinos in supersymmetry, resulting in testable predictions that entwine particle physics experiments with tests of gravity. 

This is the case for bimetric theory (BT) of gravity, or bigravity~\cite{Hassan:2011zd,Schmidt-May:2015vnx}, the only known consistent ghost-free extension of GR including a new interacting massive spin-2 field.
This theory was extensively studied in limits where the mass of the new spin-2 field is \emph{assumed} to be of order of the Hubble parameter today~\cite{Volkov:2011an,vonStrauss:2011mq,Comelli:2011zm,Berg:2012kn,Akrami:2012vf,Maeda:2013bha,Akrami:2013ffa,Aoki:2013joa}, $\mfp\sim H_0\sim10^{-33}$~eV, to model the late evolution of the Universe. Yet, once regimes relevant for particle physics are considered, this framework delivers an ideal DM candidate~\cite{Aoki:2016zgp,Babichev:2016hir,Babichev:2016bxi,Aoki:2017cnz}. Indeed, recent studies showed that a TeV-scale spin-2 particle emerging from BT is stable on cosmological scales, gravitates exactly as usual matter and can be produced in the required abundance via gravitational freeze-in~\cite{Babichev:2016hir,Babichev:2016bxi}.

Based on a direct analogy with axion-like particles (ALPs)~\cite{Raffelt:1990yz,Ringwald:2012hr,Marsh:2015xka}, 
in this work we exemplify the reach of the proposed approach by investigating the properties of light spin-2 fields from BT  in relation to the DM puzzle. The scenario we delineate is based on the so-called ``misalignment mechanism'', which sets the initial conditions for spin-2 field in cosmology. When the Hubble parameter drops below its mass, $H<\mfp$, the field starts to oscillate coherently, giving rise to a DM abundance. The oscillations are rapid enough to average out the directional anisotropies~\cite{Cembranos:2012kk,Cembranos:2012ng} imputable to the intrinsic spin.  We therefore propose a cold DM matter candidate that, similarly to ALPs, is consistent with all observations~\cite{Ade:2015xua} and that does not rely on Hubble scale anisotropies~\cite{Maeda:2013bha} nor on the existence of primordial magnetic fields~\cite{Aoki:2017cnz} for its viability.

In this Letter we show that the observed DM abundance can be matched for a wide range of spin-2 field masses, $10^{-24}~\mathrm{eV}<\mfp<{\cal O}(0.1)~\mathrm{eV}$, in agreement with the bounds imposed by the consistency of the theory and cosmological observations.  While the latter affect oscillating spin-2 fields in the same fashion as ALPs, we argue that direct detection experiments can distinguish between the two as DM couples here to the stress-energy tensors of matter and radiation. The resulting interactions with electric and magnetic fields, $E_i$ and $B_i$, are of the form $E_iE_j\pm B_iB_j$ and $E_iB_j\pm B_iE_j$, and give rise to a phenomenology different from that of the axial coupling $E\cdot B$ of ALPs\footnote{Notice that ALPs experiments not sensitive to polarisation effects, like ``shining-through-the-wall'' tests, will constrain the two scenarios in similar ways.}. On top of that, the framework we propose can be clearly distinguished from ALPs because it \emph{predicts} an oscillating electric charge due to the coupling of the spin-2 particle to the stress energy tensors of matter. The resulting directional effects can be investigated by means of atomic clocks~\cite{Arvanitaki:2014faa}, atomic spectroscopy~\cite{VanTilburg:2015oza,Hees:2016gop} and dedicated resonant mass detectors~\cite{Arvanitaki:2015iga}.

We also extend this framework to multimetric theories~\cite{Hinterbichler:2012cn}, which straightforwardly implement the clockwork mechanism~\cite{Kaplan:2015fuy,Giudice:2016yja} of particle physics and give rise to additional effects that upcoming cosmological observations might reveal.

\section*{Basics of Bimetric theory} 
\label{sec:big}

Given two spin-2 fields $f_{\mu \nu}$ and $g_{\mu \nu}$, the BT action is\footnote{Throughout this work we adopt the $(+---)$ signature for both the metrics; greek subscripts or superscripts will take values in $\{0,1,2,3\}$; latin indices span instead the restricted set $\{1,2,3\}$.}
~\cite{Hassan:2011zd}
\begin{align}
	\label{eq:action0}
	S=&
	-\frac{\mpl^2}{1+\alpha^2}\int\td^4x
	\Biggr[
	\sqrt{|g|} R(g) + \alpha^2 \sqrt{|f|} R(f) +
	\\ \nn
	&
	+ 2 
	\frac{\alpha^2\mpl^2}{1+\alpha^2} \sqrt{|g|}\,V
	\left(g,f;\beta_n\right)
	\Biggr] +  \\\nn
 	& +
	\int\td^4x\,\sqrt{|g|}\,\LG_\text{m}(g,\Psi)\,,
\end{align}
where $\mpl\approx2.4\times10^{18}$~GeV is the reduced Planck mass, and $\alpha$ is a dimensionless constant that accounts for differences in the strength of the interactions associated to the spin-2 metric fields.

On top of the usual Einstein-Hilbert kinetic terms for $\fmn$ and $\gmn$, the first line of Eq.~\eqref{eq:action0} harbours the interaction potential $V(g,f;\beta_n)$. This contains five dimensionless parameters $\beta_{n}$, $n\in[0,4]$, and is engineered around the requirement that no propagating ghost degrees of freedom appear in the theory~\cite{Boulware:1973my,deRham:2010kj,Hassan:2011zd}. The last term in Eq.~\eqref{eq:action0} is the Lagrangian of generic matter fields $\Psi$. Notice that the absence of ghost modes forces the latter to interact only with \emph{one} of the metric fields present in the action\footnote{The theory remains ghost-free also if matter couples to the same combination of the metric fields which appears in the potential. This prescription, however, leads to the same phenomenology as the case under examination.}~\cite{Yamashita:2014fga,deRham:2014naa}, effectively spoiling its symmetry under the interchange of the two metrics. For further details we refer the reader to~\cite{Schmidt-May:2015vnx}.

In order to disentangle the r\^oles of the two metrics, we linearise the theory by considering metric fluctuations around identical backgrounds $\bfmn=\bgmn$ as
\begin{align}\label{eq:metr_exp}
\gmn &= \bgmn +\epsilon \, \hmn \,,\\ \fmn &= \bgmn + \epsilon\,\lmn \,,
\end{align}
with $\epsilon$ a small expansion parameter~\cite{Babichev:2016hir,Babichev:2016bxi}.  The resulting quadratic action can be diagonalised by means of the following substitutions
\begin{align}
	\label{eq:fluct1}
\hmn &=:\frac{1}{\mpl}\left(\Gmn-\alpha\Mmn\right)\, ,\\
	\label{eq:fluct2}
\lmn &=:\frac{1}{\mpl}\left(\Gmn+\alpha^{-1}\Mmn\right)\,.
\end{align}
As we can see, the parameter $\alpha$ also quantifies the mixing between the \emph{interaction eigenstates} $\hmn$ and $\lmn$ and the \emph{mass eigenstates} $\Gmn$ and $\Mmn$.  In terms of $\Gmn$ and $\Mmn$ we find~\cite{Babichev:2016hir,Babichev:2016bxi}
\begin{align}
	\label{eq:action2}
	S^{(2)}
	:=&
	\int\td^4x\sqrt{|\bar g|}\,
	\biggr[ 	\LG^{(2)}_\text{GR}(G)+\LG^{(2)}_\text{FP}(M)
	+\\ \nn
	& 	
	-\frac{1}{\mpl}
	\left(\Gmn-\alpha\Mmn\right)T^{\mu\nu}(\Psi)\biggr]
	\,+\mathcal{O}\left(\epsilon^3\right)\,.
\end{align}
Here $\LG^{(2)}_\text{GR}(X)$ is the usual expression obtained by expanding the action of GR at the quadratic level $\LG^{(2)}_\text{GR}(X) := \mpl^2 X^{\mu\nu} \Emnlk X_{\lambda\kappa}$, with the Lichnerowicz operator defined by
\begin{align}
\Emnlk := & \delta^\lambda_\kappa \Box - \cGmn \cG^{\lambda\kappa} \Box + \cG^{\lambda\kappa} \D_\mu \D_\nu M + \\ \nn
&  +\cGmn \D^\lambda \D^\kappa - 2 \D^\lambda \D_{(\mu} \delta^\kappa_{\nu)}\,,
\end{align}
where $2X_{(\mu,\nu)} := X_{\mu\nu} + X_{\nu\mu}$.  $T^{\mu\nu}(\Psi)$ is the stress energy tensor of the matter fields $\Psi$. Indices are raised and lowered by means of the background metric $\bgmn$.

The remaining term in Eq.~\eqref{eq:action2}, $\LG^{(2)}_\text{FP}(M)$, is the quadratic Fierz-Pauli Lagrangian
\begin{equation}\label{eq:fp}
\LG^{(2)}_\text{FP}(M) := \LG^{(2)}_\text{GR}(M)-\frac{\mfp^2}{4}\left(\Mmn M^{\mu\nu} - M^2\right)\,,
\end{equation}
where we identified the Fierz-Pauli mass $\mfp := \sqrt{\beta_1 + 2\beta_2 + \beta_3}\,\mpl$ for the spin-2 field $\Mmn$.  The two remaining parameters $\beta_0$ and $\beta_4$ correspond to two identical cosmological constants that, for the purpose of this work, can be safely neglected.

Recasting Eq.~\eqref{eq:action0} in terms of the mass eigenstates $\Gmn$ and $\Mmn$ makes the field content of the theory explicit. At the linearised level, BT contains a massless spin-2 field, $\Gmn$, which possesses two helicity states likewise the usual graviton of GR, and an additional spin-2 field, $\Mmn$, characterised by a Fierz-Pauli mass $\mfp$ induced by the interaction potential $V$. Being massive, $\Mmn$ propagates five independent degrees of freedom. Notice that the coupling strength of $\Mmn$ to ordinary matter is that of the massless field times $\alpha$.

\section*{A massive spin-2 field} 
\label{sec:resum}

We define a non-linear ``background'' $\cGmn$ as
\begin{equation}\label{eq:newbkgd}
\cGmn = \bgmn+\frac{1}{\mpl}\Gmn\,.
\end{equation}
Since terms linear in $\Mmn$ and of any order in $\Gmn$ vanish in the expansion of the original action~\eqref{eq:action0}~\cite{Babichev:2016bxi,Babichev:2016hir}, we can partially re-sum the expansion to separate the effective background $\cGmn$ from the dynamics of the massive fluctuation $\Mmn$. Through this procedure we formally obtain
\begin{align}
	\label{eq:action}
S_\text{spin-2}=&-\mpl^2\int\td^4x\sqrt{|\cG|}R(\cG) + \\ \nn &+ \int\td^4x\sqrt{|\cG|}\LG^{(2)}_\text{FP}(M)\, 
+ \mathcal{O}\left(\Mmn^3\right)\,.
\end{align}
As a result,  the theory at hand contains a propagating spin-2 particle, the massive field $\Mmn$, on a generic background encoded in the metric $\cGmn$~\cite{Babichev:2016bxi,Babichev:2016hir}.

The separation of the action presented in Eq.~\eqref{eq:action} emerges naturally when $H\ll\mfp$, where $H:=\pd\log a(t)/\pd t$ is the Hubble parameter and $a(t)$ the scale factor of the Universe. The background metric $\cGmn$ has, in fact, a characteristic length and time scale of $1/H$ which is much longer than the typical wavelength $1/\mfp$ of the massive spin-2. We limit the values of $\alpha$ and $\mfp$ in a way that the theory reproduces GR within the sensitivity of gravity tests~\cite{Murata:2014nra,Will:2014kxa}: for small $\mfp\ll1$~meV we take $\alpha\ll1$, whereas we can allow $\alpha\geq1$ when $\mfp\gg1$~meV~\cite{Babichev:2013usa,Babichev:2016hir}.

The equation of motion (EOM) for the massive spin-2 field $\Mmn$ can be derived from Eq.~\eqref{eq:action} via variational derivative with respect to the field $M^{\mu\nu}$,
\begin{align}
	\label{eq:eomcov}
\Emnlk &M_{\lambda\kappa} - R\Mmn + \cGmn R^{\lambda\kappa}M_{\lambda\kappa} + \\ \nn
& + \frac12\mfp^2(\Mmn-\cGmn M) = 0\,,
\end{align}
where $M:=M_\mu^\mu$ and indices have been raised and lowered by means of the non-linear metric $\cGmn$, which also determines the expressions of the metric connection used in the covariant derivative $\nabla_\mu$ and curvature tensor.  
The structure of the EOM~\eqref{eq:eomcov} can be simplified by taking into account the linearised Bianchi identities, which imply that $\D^\mu \Mmn = \D_\nu M$, and thus $\D^\mu \D^\nu \Mmn = \Box M$.  Substituting this constraint in the trace of~\eqref{eq:eomcov} results in the tracelessness of the massive spin-2 field, $M=0$, which in turn enforces transversality: $\D^\mu \Mmn = 0$\footnote{Notice that for the massive spin-2 $\Mmn$ the transverse and traceless conditions directly descend from the Bianchi identities and are \emph{not} a gauge choice.}. These constraints ensure that the massive spin-2 field propagates only five degrees of freedom.

\section*{Coherent oscillations of a massive spin-2 field} 
\label{sec:eom}

We analyse now the dynamics of the massive spin-2 field in the early Universe, on a Friedmann-Lama\^itre-Robertson-Walker (FLRW) background,  $\cGmn = \text{diag}(1,-a^2(t),-a^2(t),-a^2(t))$, $t$ being cosmic time. Because the stress-energy tensor sourcing the FLRW background is necessarily that of a perfect fluid, it is possible to further simplify the equation of motion via the Einstein equations for the background metric and by reordering the covariant derivatives. As a result, Eq.~\eqref{eq:eomcov} reduces to
\begin{equation}
	\label{eq:eomcov1}
	\Box \Mmn + 2 R_{\mu\alpha\nu\beta} \,M^{\alpha\beta} + \mfp^2 \,\Mmn = 0\,,
\end{equation}   
where we accounted for the transversality and tracelessness of $\Mmn$. In terms of the EOM for the individual components of the massive spin-2 field, the same constraints force $M_{0\nu} = 0$ and $M^i_i = 0$ (summation implied).

The dynamics of the five remaining degrees of freedom, in Fourier space, is governed by
\begin{equation}
	 \ddot\Mij + 3H\dot\Mij + k^2\Mij + \mfp^2\Mij = 0\,,
\end{equation}
where a dot indicates differentiation with respect to $t$ (see also~\cite{Comelli:2012db, Lagos:2014lca}). Notice that these equations have been obtained from~\eqref{eq:eomcov1} without any further approximation.

With the EOM at hand, we focus on the ``homogeneous'' modes of the spin-2 field, $k\ll \mfp,$ obeying
\begin{equation}\label{eq:eomhom}
\ddot\Mij + 3H\dot\Mij + \mfp^2\Mij = 0\,.
\end{equation}
At early times, when $H\gg\mfp$, these modes are essentially massless and frozen due to Hubble friction. This regime is maintained until $H\sim\mfp$, at which point the homogenous modes begin to oscillate with a characteristic frequency $\omega=\mfp$
\begin{equation}\label{eq:sol}
\Mij \sim a(t)^{-3/2} \cos(\mfp t)\,.
\end{equation}
The resulting evolution of each component is thus identical to that of a free scalar field oscillating in the quadratic potential imposed by a mass term. 

Rapid oscillations around the minimum ($\mfp \gg H$ ) ensure that $\Mij$ behave as matter, as required of a suitable DM candidate~\cite{Marsh:2015xka,Maeda:2013bha}. This can be shown by computing the energy density $\rho_\text{DM}$ and pressure $P_\text{DM}$ from the energy-momentum tensor of $\Mmn$, defined as
\begin{equation}\label{eq:emt}
T_M^{\mu\nu} \equiv \frac{1}{\sqrt{|\cG|}}\frac{\delta\left(\sqrt{|\cG|}\,\LG^{(2)}_{\text{FP}}(M)\right)}{\delta\cGmn}\,.
\end{equation}
We find $\rho_\text{DM} \sim \dot \Mij \dot M^{ij} + \mfp^2 \Mij M^{ij}$ and $P_\text{DM} \sim \dot \Mij \dot M^{ij} - \mfp^2 \Mij M^{ij}$.  A direct computation shows that in the fast oscillating regime $\dot \Mij \dot M^{ij} = \mfp^2 \Mij M^{ij}$ and the pressure term therefore vanishes (this also descends from the virial theorem).  Hence, from the Bianchi identities,
\begin{equation}
	 \D_\mu T^{\;\;\mu}_{\!\!M\;\nu} = 0,
\end{equation}
and by assuming a perfect fluid form for $T^{\;\;\mu}_{\!\!M\;\nu} = \text{diag}(\rho_\text{M},-P_\text{M},-P_\text{M},-P_\text{M})$, we immediately infer that $\dot\rho_\text{M} + 3H\rho_\text{M} = 0$, in complete analogy with the ALPs case. The scaling of $\rho_\text{M}$ can also be computed directly by averaging the equations of motion~\eqref{eq:eomhom} in the expression for $\dot\rho_\text{M}$.

Since the oscillating massive spin-2 field behaves like matter, we estimate its contribution to the observed DM density to be~\cite{Marsh:2015xka}
\begin{equation}\label{eq:omega}
\Omega_\text{DM} \approx 2.0 \left(\frac{\mfp}{10^{-24}eV}\right)^{1/2}\left<\left(\frac{\Mij^*}{\mpl}\right)^2\right>\,,
\end{equation}
where the initial field value at the time oscillations begin\footnote{We assume that oscillations begin during the radiation domination epoch.}, $\Mij^*$, is set by the misalignment mechanism. From a phenomenological point of view, the observed DM abundance $\Omega_\text{DM}=0.26$~\cite{Ade:2015xua} 
can be matched for a wide range of $\mfp$: $10^{-24}~\mathrm{eV}\lesssim\mfp\lesssim{\cal O}(0.1)~\mathrm{eV}$. The quoted upper bound on the mass comes from the requirement of spatial coherence of the oscillating field (that is, that the field dynamics are captured by classical equations of motion), whereas the lower bound is due to the dynamics of galaxy formation (see for instance~\cite{Hlozek:2014lca,Hlozek:2016lzm}).

We remark on an important difference between oscillating scalar and spin-2 fields. Due to the intrinsic spin, the homogeneous modes of the latter are not isotropic, but must respect the stringent microwave background anisotropies bound~\cite{Ade:2013nlj}. However, similarly to the case of spin-1 field discussed in Ref.~\cite{Nelson:2011sf}, the characteristic quadrupolar anisotropy induced here is dynamically driven to negligible levels~\cite{Cembranos:2012kk,Cembranos:2012ng} via the rapid oscillations of the spin-2 field. Before oscillations begin the anisotropy is inevitably present, but it is negligible since the energy density of $\Mmn$ is vastly subdominant compared to that of other isotropic components.

\section*{Clockwork extensions of bimetric theory} 
\label{sec:clockwork}

Bimetric theory can be extended to a multimetric gravity theory by introducing extra spin-2 fields and the corresponding interaction potentials. The resulting multimetric theory is the most natural and consistent framework to implement a gravitational clockwork mechanism~\cite{Kaplan:2015fuy,Giudice:2016yja} in Nature. The basic idea behind this paradigm is to consider $N$ sites of identical physical systems, spin-2 sectors in our case, sequentially coupled to each other via interaction potentials which prevent the appearance of ghost modes. In the limit $N\to\infty$, the system then effectively acquires a continuous extra dimension. This construction aims to explain hierarchies in physics in terms of exponential dependences between physical parameters induced by the separation of the sites. In the multimetric clockwork interpretation of BT, the apparent weakness of gravity, or the unnaturally large value of the Planck scale, can be explained by the exponential suppression of a TeV-scale fundamental mass parameter appearing in a distant site of the theory which directly interacts with matter.

\section*{Phenomenology} 
\label{sec:pheno}

\subsection*{Cosmological constraints}

Large scale structure formation casts a lower bound on the oscillating field mass, $\mfp\gtrsim10^{-24}$~eV, that is roughly the inverse size of a dwarf galaxy~\cite{Hlozek:2014lca,Hlozek:2016lzm}.  Black hole superradiance, instead, excludes masses $6\times 10^{-13}~\mathrm{eV} \gtrsim \mfp\gtrsim 2\times 10^{-11}$~eV independently of the origin of the radiated field~\cite{Arvanitaki:2015iga,Arvanitaki:2016qwi}, whereas Pulsar Timing Arrays will probe in the future the lowest end of the available mass range~\cite{Khmelnitsky:2013lxt,Porayko:2014rfa}. Spin-2 masses $10^{-23}~\text{eV}\lesssim\mfp\lesssim10^{-18}$~eV can also be tested by seeking secular changes in the orbital period of binary pulsars~\cite{Blas:2016ddr} and, with different systems lying on different orbital planes, it is in principle even possible to determine the spin-2 nature of the disturbance. The lowest end of the considered mass spectrum is also constrained by the requirement of perturbativity for the cosmological solutions of BT~\cite{Babichev:2016bxi}: $\rho<\mfp^2\mpl^2/\alpha^2$, where $\rho$ is the total energy density of the Universe. By supposing a radiation dominated regime and recasting the bound in terms of the reheating temperature $T$, the requirement of successful Big Bang Nucleosynthesis, $T> {\cal O}(1)$~MeV, then implies $\mfp\gtrsim10^{-15}\sqrt{\alpha}$~eV. Clearly,  depending on the value of $\alpha$, this consistency bound might supersede the limits discussed above. As for the constraints posed by observations within the post-Newtonian formalism, we find that these scale with $\alpha$ and are in general less stringent~\cite{Hohmann:2017uxe}.

As customary in ALPs scenarios, we attributed the initial displacement of our field to a misalignment mechanism, possibly related to a gravitational phase transition resulting in the interacting potential of Eq.~\eqref{eq:action0}. Alternatively, a non-zero displacement of the field can be induced by its random Gaussian fluctuations during the de Sitter inflationary epoch. In this case we expect that~\cite{Espinosa:2015qea} $\Mij^*< 3H_\text{inf}\sqrt{N},$ where $H_\text{inf}\sim 10^{14}$~GeV is the Hubble value during inflation and $N$ is the total number of $e$-folds. This relation, together with Eq.~\eqref{eq:omega}, implies a lower bound $\mfp\gtrsim10^{-12}$~eV which greatly restricts the available parameter space. In this case, the maximal spin-2 field coherence distances is about $10^3$~km, yielding possible implications on small scale structure formation issues such as the core-vs-cusp problem \cite{deBlok:2009sp}. The inflationary origin of the spin-2 field initial displacement, however, is strongly constrained by the current limits on isocurvature perturbations\footnote{For the following discussion we assume that our oscillating spin-2 field solely accounts for the observed DM abundance.}, which the spin-2 field sources in the same way as ALPs~\cite{Marsh:2015xka}: $\Delta_\text{iso}\approx H_\text{inf}/\pi\Mij^*$. Therefore, whereas for a generic misalignment mechanism the current measurements  $\Delta_\text{ad}\sim5\times10^{-5}$~\cite{Ade:2015lrj} simply imply that $\Mij^*\gtrsim10^{-6}H_\text{inf}/\pi$, for an inflationary origin of the displacement this bound translates into a harsh constraint on the total number of $e$-fold $N\gtrsim 10^{10}$. 

As for clockwork multimetric gravity, we expect the theory to involve many spin-2 massive fields, characterised by different Fierz-Pauli masses $\mfp^{1} > \mfp^{2} > ...> \mfp^{N}$, which would start to oscillate at different times during the evolution of the Universe. As a result, depending on the initial displacement of the fields, the scenario predicts rapid changes in the energy density of the matter component of the Universe as the latter evolves. If the masses of (some of) the spin-2 fields are such that oscillations begin close to matter-radiation equality, $\mfp\sim10^{-24}$, future observations might detect the transitions eras when these fields start to contribute into the matter energy density~\cite{Amendola:2016saw}.

\subsection*{Laboratory tests}

The classical laboratory probes for new light fields are fifth force searches and tests of the equivalence principle. Both probes constrain new theories independently of their link to DM.

Fifth force experiments seek deviations from the $1/r$ scaling of the gravitational potential, induced in our case by the coupling $\alpha/\mpl$ of the massive spin-2 field to the stress-energy tensor of matter. By examining spherically symmetric solutions of BT, we predict a correction $\Delta\Phi_N = \alpha^2 \exp(-\mfp r)$ to Newton's potential~\cite{Babichev:2013usa,Babichev:2016bxi}. The most sensitive experimental results~\cite{Kapner:2006si,Murata:2014nra} imply that $\alpha \leq 10$ for $\mfp\sim10^{-1}$~eV and  $\alpha < 10^{-2}$ for $\mfp < 10^{-2}$~eV.

Tests of the equivalence principle are instead sensitive to variations of fundamental parameters of the Standard Model, such as the electron and proton mass ratio, $m_e/m_p,$ and the elementary electric charge. Although in our case the massive spin-2 field couples to proton and electron masses identically, experiments~\cite{Schlamminger:2007ht} constrain $\alpha < 10^{-1-2}$ for $10^{-24} <\mfp < 10^{-6}$~eV, covering most of the considered mass range~\cite{Arvanitaki:2015iga}. 

Equivalence principle-violating effects could be also detectable as temporal and directional variations of the elementary electric charge~\cite{Damour:2010rp}, which induce modulations in the emission lines of atoms and nuclei investigated by means of atomic clocks~\cite{Arvanitaki:2014faa}, atomic spectroscopy~\cite{VanTilburg:2015oza,Hees:2016gop} and with dedicated resonant mass detectors~\cite{Arvanitaki:2015iga}. 
The massive spin-2 field we consider, in fact, interacts with radiation according to~\cite{Han:1998sg}
\begin{equation}
	S\supset
	\frac{\alpha}{\mpl} M^{\mu\nu}\left( \frac{1}{4} \cGmn  F^{\rho\sigma}F_{\rho\sigma} - F^{\;\rho}_\mu F_{\nu\rho}\right),
	\label{FF}
\end{equation}
resulting in effective couplings to electric and magnetic field of the forms $E_iE_j\pm B_iB_j$ and $E_iB_j \pm B_iE_j$. Whereas the bounds from~\cite{Arvanitaki:2014faa,VanTilburg:2015oza,Hees:2016gop,Arvanitaki:2015iga} apply here at least as an order of magnitude estimate, the explicit form of the spin-2 coupling is clearly different from that of the scalar and pseudoscalar fields usually considered in literature, which respectively couple to electric and magnetic fields via $E^2-B^2$ and $E\cdot B$. The induced non trivial polarisation correlations and the possible directional and temporal variations of electric charge therefore constitute a distinguishing signature of the model.

Lastly, searches aimed at detecting photon-ALPs conversion in strong magnetic fields, as light-shining-through-the-wall experiments for instance, probe this scenario with the same reach as in usual ALPs model. In this case, the characteristic decay constant is to be compared to $\alpha/\mpl$.

\section*{Conclusion} 
\label{sec:con}

In this Letter we have investigated a limit of bimetric theory in which DM is explained through the coherent oscillations of a light spin-2 field. After presenting the basics of the scenario and the description of the massive spin-2 field, we computed its equation of motion in a cosmological background. By solving the latter we demonstrated that the massive homogeneous modes of the spin-2 field rapidly oscillate if $H<\mfp$, and that their stress-energy tensor matches that of matter.  Owing to the misalignment mechanism, the spin-2 field can account for the observed DM abundance across a wide range of masses $10^{-24}~\mathrm{eV}<\mfp<{\cal O}(0.1)~\mathrm{eV}$.  We also  commented on the possibilities offered by an extension to multiple metrics, which implements in a natural way the clockwork mechanism for gravity. It is fascinating that, beside traditional tests of gravitation, laboratory searches ranging from precision measurements of the fundamental electric charge with atomic clocks to dedicated resonant mass detectors, as well as experiments devoted to axion-like-particles, could contribute to unveil and comprehend the properties of gravity.

\section*{Acknowledgements} 

The authors thank Denis Comelli, Sabir Ramazanov, Tomi Koivisto, Mikael Von Strauss and Ville Vaskonen for useful discussion. Most of the presented calculations have been performed with the support of the excellent \textit{xTensor} and \textit{xCoba} \textit{Mathematica} packages~\cite{Brizuela:2008ra} developed by J.-M.~Mart\'{\i}n-Garc\'{\i}a and D.~Yllanes (\href{http://www.xact.es}{http://www.xact.es}). This research was financed by the Estonian Research Council grants IUT23-6, PUT 808 and by the European Union through the ERDF CoE grant TK133.

\bibliographystyle{hunsrt}
\bibliography{refs.bib}

\begin{thebibliography}{10}

\bibitem{Jungman:1995df}
Gerard Jungman, Marc Kamionkowski, and Kim Griest.
\newblock {Supersymmetric dark matter}.
\newblock {\em Phys. Rept.}, 267:195--373, 1996, hep-ph/9506380.

\bibitem{Bertone:2004pz}
Gianfranco Bertone, Dan Hooper, and Joseph Silk.
\newblock {Particle dark matter: Evidence, candidates and constraints}.
\newblock {\em Phys. Rept.}, 405:279--390, 2005, hep-ph/0404175.

\bibitem{Hassan:2011zd}
S.~F. Hassan and Rachel~A. Rosen.
\newblock {Bimetric Gravity from Ghost-free Massive Gravity}.
\newblock {\em JHEP}, 02:126, 2012, 1109.3515.

\bibitem{Schmidt-May:2015vnx}
Angnis Schmidt-May and Mikael von Strauss.
\newblock {Recent developments in bimetric theory}.
\newblock {\em J. Phys.}, A49(18):183001, 2016, 1512.00021.

\bibitem{Volkov:2011an}
Mikhail~S. Volkov.
\newblock {Cosmological solutions with massive gravitons in the bigravity
  theory}.
\newblock {\em JHEP}, 01:035, 2012, 1110.6153.

\bibitem{vonStrauss:2011mq}
Mikael von Strauss, Angnis Schmidt-May, Jonas Enander, Edvard Mortsell, and
  S.~F. Hassan.
\newblock {Cosmological Solutions in Bimetric Gravity and their Observational
  Tests}.
\newblock {\em JCAP}, 1203:042, 2012, 1111.1655.

\bibitem{Comelli:2011zm}
D.~Comelli, M.~Crisostomi, F.~Nesti, and L.~Pilo.
\newblock {FRW Cosmology in Ghost Free Massive Gravity}.
\newblock {\em JHEP}, 03:067, 2012, 1111.1983.
\newblock [Erratum: JHEP06,020(2012)].

\bibitem{Berg:2012kn}
Marcus Berg, Igor Buchberger, Jonas Enander, Edvard Mortsell, and Stefan Sjors.
\newblock {Growth Histories in Bimetric Massive Gravity}.
\newblock {\em JCAP}, 1212:021, 2012, 1206.3496.

\bibitem{Akrami:2012vf}
Yashar Akrami, Tomi~S. Koivisto, and Marit Sandstad.
\newblock {Accelerated expansion from ghost-free bigravity: a statistical
  analysis with improved generality}.
\newblock {\em JHEP}, 03:099, 2013, 1209.0457.

\bibitem{Maeda:2013bha}
Kei-ichi Maeda and Mikhail~S. Volkov.
\newblock {Anisotropic universes in the ghost-free bigravity}.
\newblock {\em Phys. Rev.}, D87:104009, 2013, 1302.6198.

\bibitem{Akrami:2013ffa}
Yashar Akrami, Tomi~S. Koivisto, David~F. Mota, and Marit Sandstad.
\newblock {Bimetric gravity doubly coupled to matter: theory and cosmological
  implications}.
\newblock {\em JCAP}, 1310:046, 2013, 1306.0004.

\bibitem{Aoki:2013joa}
Katsuki Aoki and Kei-ichi Maeda.
\newblock {Cosmology in ghost-free bigravity theory with twin matter fluids:
  The origin of dark matter}.
\newblock {\em Phys. Rev.}, D89(6):064051, 2014, 1312.7040.

\bibitem{Aoki:2016zgp}
Katsuki Aoki and Shinji Mukohyama.
\newblock {Massive gravitons as dark matter and gravitational waves}.
\newblock {\em Phys. Rev.}, D94(2):024001, 2016, 1604.06704.

\bibitem{Babichev:2016hir}
Eugeny Babichev, Luca Marzola, Martti Raidal, Angnis Schmidt-May, Federico
  Urban, Hardi Veerm{\"a}e, and Mikael von Strauss.
\newblock {Bigravitational origin of dark matter}.
\newblock {\em Phys. Rev.}, D94(8):084055, 2016, 1604.08564.

\bibitem{Babichev:2016bxi}
Eugeny Babichev, Luca Marzola, Martti Raidal, Angnis Schmidt-May, Federico
  Urban, Hardi Veerm{\"a}e, and Mikael von Strauss.
\newblock {Heavy spin-2 Dark Matter}.
\newblock {\em JCAP}, 1609(09):016, 2016, 1607.03497.

\bibitem{Aoki:2017cnz}
Katsuki Aoki and Kei-ichi Maeda.
\newblock {Condensate of Massive Graviton and Dark Matter}.
\newblock 2017, 1707.05003.

\bibitem{Raffelt:1990yz}
Georg~G. Raffelt.
\newblock {Astrophysical methods to constrain axions and other novel particle
  phenomena}.
\newblock {\em Phys. Rept.}, 198:1--113, 1990.

\bibitem{Ringwald:2012hr}
Andreas Ringwald.
\newblock {Exploring the Role of Axions and Other WISPs in the Dark Universe}.
\newblock {\em Phys. Dark Univ.}, 1:116--135, 2012, 1210.5081.

\bibitem{Marsh:2015xka}
David J.~E. Marsh.
\newblock {Axion Cosmology}.
\newblock {\em Phys. Rept.}, 643:1--79, 2016, 1510.07633.

\bibitem{Cembranos:2012kk}
J.~A.~R. Cembranos, C.~Hallabrin, A.~L. Maroto, and S.~J.~Nunez Jareno.
\newblock {Isotropy theorem for cosmological vector fields}.
\newblock {\em Phys. Rev.}, D86:021301, 2012, 1203.6221.

\bibitem{Cembranos:2012ng}
J.~A.~R. Cembranos, A.~L. Maroto, and S.~J. N{\'u}{\~n}ez~Jare{\~n}o.
\newblock {Isotropy theorem for cosmological Yang-Mills theories}.
\newblock {\em Phys. Rev.}, D87(4):043523, 2013, 1212.3201.

\bibitem{Ade:2015xua}
P.~A.~R. Ade et~al.
\newblock {Planck 2015 results. XIII. Cosmological parameters}.
\newblock {\em Astron. Astrophys.}, 594:A13, 2016, 1502.01589.

\bibitem{Arvanitaki:2014faa}
Asimina Arvanitaki, Junwu Huang, and Ken Van~Tilburg.
\newblock {Searching for dilaton dark matter with atomic clocks}.
\newblock {\em Phys. Rev.}, D91(1):015015, 2015, 1405.2925.

\bibitem{VanTilburg:2015oza}
Ken Van~Tilburg, Nathan Leefer, Lykourgos Bougas, and Dmitry Budker.
\newblock {Search for ultralight scalar dark matter with atomic spectroscopy}.
\newblock {\em Phys. Rev. Lett.}, 115(1):011802, 2015, 1503.06886.

\bibitem{Hees:2016gop}
A.~Hees, J.~Gu{\'e}na, M.~Abgrall, S.~Bize, and P.~Wolf.
\newblock {Searching for an oscillating massive scalar field as a dark matter
  candidate using atomic hyperfine frequency comparisons}.
\newblock {\em Phys. Rev. Lett.}, 117(6):061301, 2016, 1604.08514.

\bibitem{Arvanitaki:2015iga}
Asimina Arvanitaki, Savas Dimopoulos, and Ken Van~Tilburg.
\newblock {Sound of Dark Matter: Searching for Light Scalars with Resonant-Mass
  Detectors}.
\newblock {\em Phys. Rev. Lett.}, 116(3):031102, 2016, 1508.01798.

\bibitem{Hinterbichler:2012cn}
Kurt Hinterbichler and Rachel~A. Rosen.
\newblock {Interacting Spin-2 Fields}.
\newblock {\em JHEP}, 07:047, 2012, 1203.5783.

\bibitem{Kaplan:2015fuy}
David~E. Kaplan and Riccardo Rattazzi.
\newblock {Large field excursions and approximate discrete symmetries from a
  clockwork axion}.
\newblock {\em Phys. Rev.}, D93(8):085007, 2016, 1511.01827.

\bibitem{Giudice:2016yja}
Gian~F. Giudice and Matthew McCullough.
\newblock {A Clockwork Theory}.
\newblock {\em JHEP}, 02:036, 2017, 1610.07962.

\bibitem{Boulware:1973my}
D.~G. Boulware and Stanley Deser.
\newblock {Can gravitation have a finite range?}
\newblock {\em Phys. Rev.}, D6:3368--3382, 1972.

\bibitem{deRham:2010kj}
Claudia de~Rham, Gregory Gabadadze, and Andrew~J. Tolley.
\newblock {Resummation of Massive Gravity}.
\newblock {\em Phys. Rev. Lett.}, 106:231101, 2011, 1011.1232.

\bibitem{Yamashita:2014fga}
Yasuho Yamashita, Antonio De~Felice, and Takahiro Tanaka.
\newblock {Appearance of Boulware--Deser ghost in bigravity with doubly coupled
  matter}.
\newblock {\em Int. J. Mod. Phys.}, D23:1443003, 2014, 1408.0487.

\bibitem{deRham:2014naa}
Claudia de~Rham, Lavinia Heisenberg, and Raquel~H. Ribeiro.
\newblock {On couplings to matter in massive (bi-)gravity}.
\newblock {\em Class. Quant. Grav.}, 32:035022, 2015, 1408.1678.

\bibitem{Murata:2014nra}
Jiro Murata and Saki Tanaka.
\newblock {A review of short-range gravity experiments in the LHC era}.
\newblock {\em Class. Quant. Grav.}, 32(3):033001, 2015, 1408.3588.

\bibitem{Will:2014kxa}
Clifford~M. Will.
\newblock {The Confrontation between General Relativity and Experiment}.
\newblock {\em Living Rev. Rel.}, 17:4, 2014, 1403.7377.

\bibitem{Babichev:2013usa}
Eugeny Babichev and C{\'e}dric Deffayet.
\newblock {An introduction to the Vainshtein mechanism}.
\newblock {\em Class. Quant. Grav.}, 30:184001, 2013, 1304.7240.

\bibitem{Comelli:2012db}
D.~Comelli, M.~Crisostomi, and L.~Pilo.
\newblock {Perturbations in Massive Gravity Cosmology}.
\newblock {\em JHEP}, 06:085, 2012, 1202.1986.

\bibitem{Lagos:2014lca}
Macarena Lagos and Pedro~G. Ferreira.
\newblock {Cosmological perturbations in massive bigravity}.
\newblock {\em JCAP}, 1412:026, 2014, 1410.0207.

\bibitem{Hlozek:2014lca}
Ren{\'e}e Hlozek, Daniel Grin, David J.~E. Marsh, and Pedro~G. Ferreira.
\newblock {A search for ultralight axions using precision cosmological data}.
\newblock {\em Phys. Rev.}, D91(10):103512, 2015, 1410.2896.

\bibitem{Hlozek:2016lzm}
Ren{\'e}e Hlo{\v z}ek, David J.~E. Marsh, Daniel Grin, Rupert Allison,
  Jo~Dunkley, and Erminia Calabrese.
\newblock {Future CMB tests of dark matter: Ultralight axions and massive
  neutrinos}.
\newblock {\em Phys. Rev.}, D95(12):123511, 2017, 1607.08208.

\bibitem{Ade:2013nlj}
P.~A.~R. Ade et~al.
\newblock {Planck 2013 results. XXIII. Isotropy and statistics of the CMB}.
\newblock {\em Astron. Astrophys.}, 571:A23, 2014, 1303.5083.

\bibitem{Nelson:2011sf}
Ann~E. Nelson and Jakub Scholtz.
\newblock {Dark Light, Dark Matter and the Misalignment Mechanism}.
\newblock {\em Phys. Rev.}, D84:103501, 2011, 1105.2812.

\bibitem{Arvanitaki:2016qwi}
Asimina Arvanitaki, Masha Baryakhtar, Savas Dimopoulos, Sergei Dubovsky, and
  Robert Lasenby.
\newblock {Black Hole Mergers and the QCD Axion at Advanced LIGO}.
\newblock {\em Phys. Rev.}, D95(4):043001, 2017, 1604.03958.

\bibitem{Khmelnitsky:2013lxt}
Andrei Khmelnitsky and Valery Rubakov.
\newblock {Pulsar timing signal from ultralight scalar dark matter}.
\newblock {\em JCAP}, 1402:019, 2014, 1309.5888.

\bibitem{Porayko:2014rfa}
N.~K. Porayko and K.~A. Postnov.
\newblock {Constraints on ultralight scalar dark matter from pulsar timing}.
\newblock {\em Phys. Rev.}, D90(6):062008, 2014, 1408.4670.

\bibitem{Blas:2016ddr}
Diego Blas, Diana~Lopez Nacir, and Sergey Sibiryakov.
\newblock {Ultralight Dark Matter Resonates with Binary Pulsars}.
\newblock {\em Phys. Rev. Lett.}, 118(26):261102, 2017, 1612.06789.

\bibitem{Hohmann:2017uxe}
Manuel Hohmann.
\newblock {Post-Newtonian parameter γ and the deflection of light in
  ghost-free massive bimetric gravity}.
\newblock {\em Phys. Rev.}, D95(12):124049, 2017, 1701.07700.

\bibitem{Espinosa:2015qea}
Jose~R. Espinosa, Gian~F. Giudice, Enrico Morgante, Antonio Riotto, Leonardo
  Senatore, Alessandro Strumia, and Nikolaos Tetradis.
\newblock {The cosmological Higgstory of the vacuum instability}.
\newblock {\em JHEP}, 09:174, 2015, 1505.04825.

\bibitem{deBlok:2009sp}
W.~J.~G. de~Blok.
\newblock {The Core-Cusp Problem}.
\newblock {\em Adv. Astron.}, 2010:789293, 2010, 0910.3538.

\bibitem{Ade:2015lrj}
P.~A.~R. Ade et~al.
\newblock {Planck 2015 results. XX. Constraints on inflation}.
\newblock {\em Astron. Astrophys.}, 594:A20, 2016, 1502.02114.

\bibitem{Amendola:2016saw}
Luca Amendola et~al.
\newblock {Cosmology and Fundamental Physics with the Euclid Satellite}.
\newblock 2016, 1606.00180.

\bibitem{Kapner:2006si}
D.~J. Kapner, T.~S. Cook, E.~G. Adelberger, J.~H. Gundlach, Blayne~R. Heckel,
  C.~D. Hoyle, and H.~E. Swanson.
\newblock {Tests of the gravitational inverse-square law below the dark-energy
  length scale}.
\newblock {\em Phys. Rev. Lett.}, 98:021101, 2007, hep-ph/0611184.

\bibitem{Schlamminger:2007ht}
Stephan Schlamminger, K.~Y. Choi, T.~A. Wagner, J.~H. Gundlach, and E.~G.
  Adelberger.
\newblock {Test of the equivalence principle using a rotating torsion balance}.
\newblock {\em Phys. Rev. Lett.}, 100:041101, 2008, 0712.0607.

\bibitem{Damour:2010rp}
Thibault Damour and John~F. Donoghue.
\newblock {Equivalence Principle Violations and Couplings of a Light Dilaton}.
\newblock {\em Phys. Rev.}, D82:084033, 2010, 1007.2792.

\bibitem{Han:1998sg}
Tao Han, Joseph~D. Lykken, and Ren-Jie Zhang.
\newblock {On Kaluza-Klein states from large extra dimensions}.
\newblock {\em Phys. Rev.}, D59:105006, 1999, hep-ph/9811350.

\bibitem{Brizuela:2008ra}
David Brizuela, Jose~M. Martin-Garcia, and Guillermo~A. Mena~Marugan.
\newblock {xPert: Computer algebra for metric perturbation theory}.
\newblock {\em Gen. Rel. Grav.}, 41:2415--2431, 2009, 0807.0824.

\end{thebibliography}

\end{document}